\documentstyle[prc,preprint,tighten,aps]{revtex}
\begin {document}
\draft
\title{
SMMC studies of $N=Z$
$pf$-shell nuclei with pairing-plus-quadrupole Hamiltonian}
\author{K. Langanke$^{1}$, P. Vogel$^{1,2}$, and
Dao-Chen Zheng$^3$}
\address{
$^1$ Institute for Physics and Astronomy, Aarhus University, Aarhus, Denmark\\
$^2$Physics Division, California Institute of Technology, Pasadena, CA
91125, USA\\ 
$^3$W.~K. Kellogg Radiation Laboratory, California 
Institute of Technology, Pasadena, California 91125\\
}
\date{\today}

\maketitle

\begin{abstract}
\noindent
We perform
Shell Model Monte Carlo calculations 
of selected $N=Z$ $pf$-shell nuclei with a schematic 
hamiltonian
containing  isovector pairing and  quadrupole-quadrupole
interactions. Compared to realistic interactions, this hamiltonian does
not give rise to the SMMC ``sign problem'', while
at the same time resembles essential
features of the realistic interactions. We study pairing correlations in
the ground states of $N=Z$ nuclei and
investigate the thermal dependence of selected observables
for the odd-odd nucleus $^{54}$Co and the even-even nuclei
$^{60}$Zn and $^{60}$Ni. 
Comparison of the present results to those with the realistic KB3
interaction indicates a transition 
with increasing temperature
from a phase of isovector
pairing dominance to one where isoscalar
pairing correlations dominate. 
In addition, our results confirm the qualitative reliability of the
procedure used to cure the sign problem in the SMMC calculations
with realistic forces.
\end{abstract}

\narrowtext

\section{Introduction}

Motivated by fundamental nuclear structure questions and by
astrophysical applications, the study of extremely neutron- and
proton-rich nuclei is currently at the forefront of research interests
in nuclear physics. Novel 
proton-rich radioactive ion-beam facilities offer the possibility
of exploring the structure
of nearly self-conjugate ($N \sim Z$) nuclei in the medium mass range 
$Z \lesssim 50$. i
A focus of  interest will be 
the proton-neutron
($pn$) interaction, which 
has long been recognized to play
an important role in $N=Z$ nuclei 
(see Ref. \cite{Goodman} for an early review).
Of particular relevance
in self-conjugate odd-odd nuclei in the $pf$ shell 
should be the  $pn$ isovector correlations 
as it is evident from the ground state spins
and isospins. While the $sd$-shell odd-odd $N=Z$ nuclei
(with the exception of $^{34}$Cl)
have ground states with isospin $T=0$ and angular momenta $J > 0$,
self-conjugate odd-odd $N=Z$ nuclei in the
$pf$ shell ($A > 40$)  have ground states with $T=1$
and $J^{\pi}=0^+$ (the only
known exception is $^{58}$Cu) indicating the dominance of isovector
$pn$ pairing.
In addition,
in $^{74}$Rb the ground state isospin is also
$T = 1$ \cite{Rudolph}, again arising from isovector 
$pn$ correlations.

Early studies of isovector and isoscalar
$pn$ pairing for $sd$-shell nuclei
\cite{Sandhu} and for nuclei at the beginning of the $pf$-shell \cite{Wolter}
used the Hartree-Fock-Bogoliubov (HFB) formalism. 
Their conclusion \cite{Goodman} that the $T=1$ $pn$ pairing 
is unimportant is surprising since, as pointed out above,
the $T=1$ ground state isospin
of most odd-odd $N=Z$ nuclei with $A\ge40$ clearly points to
the role of $T=1$ $pn$ pairing in these nuclei.

Although HFB calculations have pioneered the study of
pairing in $N=Z$ nuclei,
the method of choice to study pair correlations
is the interacting shell model.
Within the $sd$ shell \cite{Wildenthal} and at the beginning of the
$pf$-shell \cite{McGrory,Caurier}
the interacting shell model has proven to give an excellent
description of all nuclei, including the correct reproduction
of the spin-isospin assignments of self-conjugate $N=Z$ nuclei.
However, the conventional shell model using diagonalization techniques
is currently restricted to nuclei with masses $A \leq50$ due to computational
limitations. These limitations are overcome by
the Shell Model Monte Carlo (SMMC) 
approach \cite{Johnson,Lang}. Using this novel method,
it has been demonstrated \cite{Langanke1} that complete $pf$ shell 
calculations using the modified Kuo-Brown interaction well reproduce the
ground state properties of even-even  $N=Z$ nuclei with $A \leq 60$. 
Additionally the SMMC approach naturally allows the study of thermal
properties. 

In this paper we use  the SMMC method. However, 
instead of the realistic 
nucleon-nucleon interaction we use a simplified interaction
containing only the isovector pairing interaction and
the quadrupole-quadrupole force. Such interaction is
nevertheless able to reproduce semi-quantitatively
the essential features of 
nuclear structure such as the pairing gaps and B$(E2)$ values
\cite{Bes}. At the same time such a hamiltonian does not
give rise to the ``sign problem'' which with the realistic
interaction can be overcome only by the ``g-extrapolation''
\cite{Dean}.
The present calculations, therefore, serve a dual purpose.
On one hand, they represent a test of the ``g-extrapolation''
as far as main nuclear properties are concerned. At the same
time, they allow us to draw important conclusion with much more
modest computational effort and without inducing potential systematic
errors.

\section{Model}

The SMMC approach was developed in Refs.
\cite{Johnson,Lang}, where the reader can find a detailed
description of the ideas underlying the method, its formulation,
and numerical realization. As the present calculations follow
the formalism developed and published previously, a very brief
description of the SMMC approach suffices here. 
A comprehensive review of the SMMC method and its 
applications can be found in
Ref. \cite{report}.

The SMMC method
describes the nucleus by a canonical ensemble at temperature  
$T=\beta^{-1}$ and employs a Hubbard-Stratonovich linearization 
\cite{Hubbard} of the  
imaginary-time many-body propagator, $e^{-\beta H}$, to express  
observables as path integrals of one-body propagators in fluctuating  
auxiliary fields \cite{Johnson,Lang}. Since Monte Carlo techniques  
avoid an explicit enumeration of the many-body states, they can be  
used in model spaces far larger than those accessible to conventional  
methods. The results are in principle exact and are in  
practice subject only to controllable sampling and discretization  
errors. 
However, SMMC studies with ``realistic interactions'' are hampered by
potential systematic uncertainties introduced by the infamous
sign-problem \cite{Alhassid}. These problems are avoided, and the SMMC
is in fact an exact solution to the many-body shell model problem, if a
Hamiltonian of the form pairing+multipole-multipole
interaction is used \cite{Lang}. 
The schematic hamiltonian we use in the present work is of this form
\begin{equation}
H = \sum_{jmt_z}\epsilon(j)a_{jmt_z}^\dagger a_{jmt_z} - 
G/4 \sum_{jj't_z} A_{JM=00}^{T=1t_z \dagger} 
A_{JM=00}^{T=1t_z}
-\chi\sum_{\mu} (-1)^{\mu}Q_{\mu}Q_{-\mu} ~,
\end{equation}
where $Q_{\mu}$ is the mass quadrupole moment operator with projection
$\mu$, $ a_{jmt_z}^\dagger$ creates a nucleon of isospin 
projection $t_z$ in the orbital $jm$, 
and the two particle creation operator $A^{\dagger}$
is defined below.

The summation over the single particle energies is restricted to the
four states in the $pf$ shell. The single particle energies are 
taken from the original Kuo-Brown interaction KB3 \cite{KB3}.
We use the pairing constant $G=20/A$ MeV from \cite{Bes}.
In calculating the ground state properties we cool the nuclei
to $T=0.25$ MeV. Experience shows that this is usually sufficient
to have only minimal thermal admixtures of excited states.

In order to measure
the overall pair correlations
in nuclear wave functions, we use
the BCS pair operator (for $T = 1, JM = 00$)
\begin{equation}
\Delta_{JM}^{Tt_z ~ \dagger} = \sum_{\alpha}  A_{JM}^{ Tt_z \dagger}
(\alpha) ~,~ A_{JM}^{ Tt_z \dagger }(\alpha) =
[a_a^\dagger a_b^\dagger]_{JM}^{Tt_z} ,
\end{equation}
where $\alpha$ is an index combining the single particle labels $a$ and
$b$. The quantity 
\begin{equation}
{\cal N}_{t_z} = 
\sum_M \langle \Delta_{JM}^{Tt_z\dagger} \Delta_{JM}^{Tt_z} \rangle
\end{equation}
is then a measure of the 
strength of pair correlations with isospin projection $t_z$,
or in other words, of the number of nucleon pairs with the angular
momentum $J$ and isospin and its projection $Tt_z$.
In the following we will restrict ourselves to isovector s-wave pairing
($J=0$), which, however, plays the most important role in the low-energy
spectrum of the nuclei of interest in this work.
Earlier SMMC results for BCS pairing in nuclei in the mass range
$A=48-60$ obtained with the ``realistic'' KB3 interaction \cite{KB3}
and involving therefore the g-extrapolation
are published in Refs. \cite{Langanke1,report,Langanke2,Engel}.

We still have to fix the coupling constant $\chi$ of the
quadrupole force.
Compared with the original parametrization
$\chi = 240/(b^4 A^{5/3})$ MeV/fm$^4$ ($b$ is the oscillator length
unit) as
suggested in Ref. \cite{Bes} we have
rescaled the
strength of the quadrupole-quadrupole interaction by a factor 0.6.
The rescaled interaction then gives approximately the same results
for $\langle Q_p^2 \rangle$, $\langle Q_n^2 \rangle$, and
$\langle (Q_p+Q_n)^2 \rangle$ ($Q_{p(n)}$ is the proton (neutron)
quadrupole operator) as the realistic KB3 interaction
\cite{KB3}.
The latter has been demonstrated to give a good description of $B(E2)$
strength in the mass range $A=48-60$ \cite{Caurier,Langanke1}.
Unlike in the original application of the pairing plus quadrupole
hamiltonian \cite{Bes}, we use now a
pairing interaction which is isospin symmetric, i.e. we have
replaced the like-particle-only pairing
($nn$ and $pp$) by the general isovector one.

\section{Results}

In Fig. 1 we show the BCS pairing strengths ${\cal N}_{t_z}$,
eq. (3), for $pp = nn$ ($t_z = \pm 1$) and
$pn$ ($t_z$ = 0) correlations in the ground states of $N=Z$ nuclei.
Generally  the schematic hamiltonian yields 
similar but slightly larger isovector
correlations than the realistic hamiltonian (the results for
that interaction are taken from \cite{Langanke2,Engel})
as also seen in Fig. 1. 

As a striking feature Fig. 1 shows a strong staggering in both the $pp$
and $pn$ pairing strengths when comparing neighboring even-even and
odd-odd self-conjugate nuclei. In the latter, the isovector $pn$ pairing
clearly dominates the $pp$ (and the identical $nn$) pairing and is
always significantly larger than in the neighboring even-even $N=Z$
nuclei. In contrast, the like-nucleon pairing is noticeably reduced in
the odd-odd nuclei relative to the values in the neighboring even-even
nuclei. As pointed out by Engel et al. \cite{Engel}
the increased strength in $pn$ pairing caused by the extra isovector
proton-neutron pair appears to be a salient feature in odd-odd $N=Z$
nuclei with ground state isospin $T=1$. 

As an added bonus,
we also note that the good agreement of the  SMMC pairing results with
the realistic KB3 interaction and with the schematic hamiltonian gives
confidence
in the g-extrapolation required in the SMMC calculations with the KB3
interaction due to the ``sign problem''. 

As is obvious from Fig. 1, and has already been stressed elsewhere
(e.g. \cite{Langanke2,Engel}), odd-odd $N=Z$ nuclei 
are the ideal place to study isovector $pn$ pairing
correlations. We have used the self-conjugate odd-odd nucleus
$^{58}$Cu as a first  example to explore some of the physics present in our
model.  
To develop an understanding of the relative importance of the two parts
of the interaction on the results, we have calculated several observables
as a function of the
pairing strength where we have scaled the coupling constant
$G$ of the pairing
interaction in the hamiltonian by a factor $\lambda$. 
In Fig. 2 we show how the $pn$ and $pp$ pairing depends on the
quantity $\lambda G$.
Obviously both $pp$ and $pn$ correlations
increase with increasing $\lambda$, reaching saturation at
about $\lambda = 3$. For all positive $\lambda$
isovector $pn$
correlations clearly dominate over $pp$ or $nn$ pairing. 
At $\lambda=0$ the pairing correlations, using our definition, 
do not vanish since they get a contribution from the 
mean field and from the quadrupole force.
Associated with the change in pair correlations,
the isospin expectation value, shown in Fig. 3, increases rapidly from
$\langle T^2 \rangle =0.58 $ at $\lambda=0$ to  
$\langle T^2 \rangle \approx 2 $  for $\lambda \ge 1$. In the same
$\lambda$ interval the angular momentum, also shown in Fig. 3,
drops from $\langle J^2 \rangle =8$ to 
$\langle J^2 \rangle =2$  (The angular momentum
is not strictly zero even for large $\lambda$
since at the finite temperature $T=0.25$ MeV,
used in the present calculation in lieu of $T=0$, excited
states with 
$\langle J^2 \rangle >0$  
are slightly mixed into the expectation values).

The importance of the quadrupole interaction on the results are only
minor. To see this, the results at $\lambda=0$ 
in Figs. 2-3 should be compared with
the ones without interaction (which we will call ``mean-field results").
We find 
$\langle J^2 \rangle =8.5$,  
$\langle T^2 \rangle =0.8$,  ${\cal N}_0 =4.1$ and 
${\cal N}_1 =4.0$ on the
mean-field level. Thus the quadrupole interaction has only 
noticeable influence on the isospin which it shifts
towards $T=0$.

In Fig. 4 we show the ground state
energy expectation values for the $N=Z$ nuclei. 
The odd-odd nuclei are slightly less bound than
the neighboring even-even nuclei. The displacement is approximately
1.1 MeV, and shows that the model with isovector pairing nevertheless
accounts for the usual pairing displacement between the odd-odd and
even-even nuclei.
However, the calculated displacement is only about half of the
experimental one. The discrepancy reflects the schematic nature
of our hamiltonian; it might be related to the lack of
isoscalar $pn$ interaction.

Engel $et~al.$ \cite{Engel} have shown that within an isotope chain 
with neutron (or proton) excess
proton-neutron pairing is reduced, while the pairing among protons and
among neutrons is increased. This result agrees with our findings, as is
demonstrated in Fig. 5 using the nickel isotopes as an example. 
In the even-even $N=Z$ nucleus $^{56}$Ni with neutron
number $N=28$ $pp$, $nn$, and $pn$ pairing is identical. Clearly $pn$
correlations drop drastically with either proton or neutron excess,
while $pp$ and $nn$ pairing increases. The turn-over in $nn$ pairing for
$N<26$ is simply caused by the approach to the empty neutron shell at
$N=20$, which also affects the $pn$ pairing.
We also show in Fig. 5 
the average pairing (i.e. one third of the total pairing
energy) for the series of nickel isotopes. Unlike its components,
the total pairing energy is quite smooth, it depends only weakly
on $N - Z$.

We now turn to the discussion of the thermal dependence of various
observables.
In Refs. \cite{Langanke1,Alhassid,Langanke2} SMMC studies of the
thermal dependence of selected observables have been presented for
several even-even nuclei (including the $N=Z$ nucleus
$^{52}$Fe) and the odd-odd $N=Z$ nucleus $^{50}$Mn. All
these calculations had been performed with the KB3 interaction.  
As expected, a large excess of $J=0^+$ like-particle pairing had been
found in the ground states of the even-even nuclei with neutron excess.
With increasing temperature, these pairing correlations decrease and at
around $T=1$ MeV the like-particle pairs break in these nuclei. In 
the even-even $N=Z$ nucleus $^{52}$Fe an (approximate) isospin symmetry
results in nearly identical $pp$, $nn$, and $pn$ pairing; the pairs in
all three isovector channels break at around $T=1$ MeV.
The thermal dependence of odd-odd 
$N=Z$ nuclei is different.
The $pn$
correlations, which dominate the ground state of 
the odd-odd $N = Z$ nuclei, decrease
rapidly with temperature,
while the like-particle pairing remains
roughly constant to $T \approx$1.1 MeV. 

In order to elucidate these features further,
we have performed SMMC calculations with the schematic hamiltonian for
the even-even $N=Z$ nucleus $^{60}$Zn and for the odd-odd $N=Z$ nucleus
$^{54}$Co. These studies are aimed at verifying the g-extrapolation
required in the previous SMMC studies and at testing which of the
essential physics observed in the temperature dependence of the pair
correlations and of selected observables is already reproduced by the
schematic pairing plus quadrupole hamiltonian. The SMMC calculations have
been performed for the temperature range 0.25 MeV to 5 MeV.

In Figs. 6-9 we show the temperature dependence of the expectation
values of the energy, the isovector pair correlations, the 
isospin and the angular momentum for $^{54}$Co and
$^{60}$Zn. The energy expectation value 
$\langle H \rangle$, shown in Fig. 6, 
increases, as required by general thermodynamic principles, with
temperature, and it does so roughly linearly in the interval $T=0.5-2$
MeV. For the even-even $^{60}$Zn, like in $^{52}$Fe studied
earlier with the KB3 interaction, the energy expectation
value increases very slowly at low temperatures ($T < 0.5$ MeV).
This low-temperature behavior is well known 
and is related to the
fact that in even-even nuclei the isovector pairing gap has to be
overcome. Correspondingly, the heat capacity $C(T)=d\langle H
\rangle/dT$ shows a
significant excess over the mean-field values for temperatures
$T=0.6-1.4$ MeV, with the largest excess at around $T=1$
MeV where the $J=0^+$ pairs break.
At higher temperatures $C(T)$ becomes negative 
in both nuclei as
the finite model space requires $\langle H \rangle$ to approach a
constant value in the high temperature limit. 

The pair correlations are shown in two panels in Fig. 7.
Like in the calculation with the KB3 interaction, the proton-neutron
correlations in the odd-odd $^{54}$Co
decrease very rapidly with temperature for $T<1$ MeV, 
while the like-particle correlations remain about 
constant in this temperature
interval.  At higher temperatures, $T>1$ MeV, 
the isovector correlations slowly vanish, but
remain larger than the mean-field values even up to temperatures $T=5$
MeV. The behavior of pairing correlations in the even-even nucleus $^{60}$Zn
is rather different. The $pn$ and like-particle correlations 
remain essentially identical at all temperatures.

Can one understand why the thermal dependence of the pairing strength 
is so different in the odd-odd and even-even $N = Z$ nuclei?
The difference is explained by the uniqueness of the isospin
properties of the odd-odd $N = Z$ nuclei. These nuclei are the
only ones where states of different isospin, $T = 1$ and $T=0$,
are found close to each other at low excitation energies. 
In $^{54}$Co the ground state is $T=1$ with $T_Z = 0$
and $J^{\pi} = 0^+$.
As explained above in that state the $pn$ pair correlations
dominate, and the like-particle correlations are reduced.
However, at relatively low excitation energy
one finds in these nuclei
a multiplet of $T=0$ states. Such states have necessarily
nonvanishing angular momenta, and thus contribute efficiently
to the corresponding thermal average. On the other hand, from 
isospin symmetry it follows that in the $T=0$ states all
three pairing strength ${\cal N}_{t_z}$ must be equal. Hence,
at temperatures where the $T = 0$ states dominate the thermal average,
the $pn$ pair correlations are substantially reduced when
compared to their ground state values, and the like-particle
pairing correlations are somewhat enhanced, as seen in Fig. 7.
This behavior is not restricted to the cases studied; it is quite
generic and should be present in all odd-odd $N=Z$ nuclei with
ground state isospin $T=1$. In the $sd$ shell the $T=1$ state is usually
an excited state at a low excitation energy. Thus its weight in a
thermal average is strongly reduced compared to the $pf$ shell nuclei.
Consequently, there will be no dominance of $pn$ correlations at low
temperatures in odd-odd $N=Z$ nuclei in the $sd$ shell.

The dependence of the isospin expectation value $\langle T^2 \rangle$
on temperature is also different in the odd-odd and even-even
nuclei as seen in Fig. 8. While the isospin steadily increases
in the even-even $^{60}$Zn, it decreases first from its initial
value of 2 (corresponding to $T = 1$) as the low-lying isospin $T=0$
states become populated. The effect of the isovector pairing is clearly
visible in a comparison with the mean-field values (obtained with the
two-body interaction switched off). At low temperatures $T \leq 0.5$ MeV
the isovector pair condensate results in the
total isospin $T=1$ in the odd-odd
nucleus, while in the even-even
the isovector condensate makes total isospin $T=0$.
At higher temperatures ($T > 1.5$ MeV)
both nuclei behave more or less similarly.
An SMMC calculation with the KB3
interaction showed that isoscalar correlations, missing in our schematic
Hamiltonian, reduce the isospin to nearly $\langle T^2 \rangle =0$ at
temperatures around 1 MeV in odd-odd nuclei. 

The angular momentum expectation value $\langle J^2 \rangle$,
shown in Fig. 9, increases
rapidly with temperature from $\langle J^2 \rangle =0$ at $T<0.25$ MeV.
Again, the presence of states with both isospins $T$ = 1 and 0 in
$^{54}$Co and the pairing gap in $^{60}$Zn
causes more rapid increase of angular momentum
in the odd-odd $^{54}$Co than in the even-even $^{60}$Zn.

Another aspect of the same phenomenon is the behavior of
the level density parameter 
\begin{equation}
a = \frac{{\rm d}\langle H \rangle }{{\rm d} T} /2T ~.
\end{equation} 
The rapid decrease in $pn$ correlations in
the odd-odd nucleus generates an excess of levels,
i.e. an increase in $a$ (over the mean field) at lower
temperatures than in  the even-even nucleus where there is 
a strong pairing gap in all three
isovector pairing channels.
At higher temperatures, when the isovector pairs break,
the even-even nuclei experience therefore
more noticeable increase of the level density, with
an increase in $a$ over the mean-field value of about 1.5/MeV for $^{60}$Zn 
and of about 1/MeV in $^{54}$Co at  $T=1$ MeV.
This behavior of the level density
is therefore linked to the difference in the way the
isovector $pn$ and the
like-particle correlations behave in these two systems as
discussed above. 

For realistic interactions the sign-problem is reduced with
increasing temperature making direct SMMC calculations feasible without
invoking the ``g-extrapolation'' procedure. We have performed such SMMC
calculations for the KB3 interaction at $T \geq 2.5$ MeV. 
The results obtained for the isovector pair correlations and for
$\langle J^2 \rangle$ are identical to those of our schematic
Hamiltonian, confirming again that this interaction indeed describes
these quantities very well.
However, for both nuclei $^{54}$Co and
$^{60}$Zn the isospin expectation values are significantly lower than
for the schematic Hamiltonian (see Fig. 8). 
The origin of the different behavior are the  isoscalar
$pn$ correlations, which are missing in the schematic Hamiltonian. 
As has been shown previously, the isoscalar $pn$ 
correlations decrease less rapidly with increasing 
temperature than the isovector
correlations and compete with the latter at moderate temperatures (say $T
\geq 1.5$ MeV). These isoscalar correlations lower the energies of
states with isospin $T=0$ compared to calculations where these
correlations are missing. Hence the isospin expectation value is smaller
at moderate temperatures if the isoscalar correlations are included.
To estimate the importance of isovector versus isoscalar correlations at
these high temperatures we refer to the two following observations.
First, by comparing $\langle T^2 \rangle$ for the realistic KB3 interaction 
and for the schematic Hamiltonian we find a reduction by more than a
factor of two due to isoscalar correlations, while the schematic
Hamiltonian gives results which are only about $10 \%$ lower than the
mean-field values. Second, the isospin expectation values calculated
with the KB3 interaction are nearly the same for the odd-odd and
even-even $N=Z$ nuclei, in contrast to the marked differences between
odd-odd and even-even $N=Z$ nuclei at low temperatures induced by the
dominating isovector pairing, as discussed above.
We conclude that at moderate temperatures the $N = Z$
nuclei are dominated by isoscalar correlations and that there is a
transition from isovector to isoscalar dominance at lower temperatures.
These findings are in agreement
with the
conclusions drawn in a previous SMMC calculation \cite{Langanke2}.
However, we also like to mention that the g-extrapolation invoked in
previous SMMC studies with the KB3 interaction leads to a
slight underestimation of the isovector pair correlations 
(by about $20 \%$ at $T=2.5$ MeV) as the linear dependence on $g$, found
for $g \leq 0$, i.e. for increased isovector pairing, does not 
hold for $0 \leq g \leq 1$.
 
To study the influence of neutron excess on the pairing properties we
have finally performed SMMC calculations of $^{60}$Ni as a function of
temperature. $^{60}$Ni differs from its isobar $^{60}$Zn by
having an extra neutron pair instead of a proton one.
The energy expectation value
shows the temperature dependence typical for even-even nuclei, similar
to $^{60}$Zn (Fig. 10). However, the pair correlations,
shown in Fig. 11, are rather
different than those in $^{60}$Zn (see Fig. 7). 
Both like-nucleon correlations
show strong excesses at low temperatures over the mean-field values. At
$T=1$ MeV about half of the excess has vanished, as in other SMMC
calculations of even-even nuclei in this mass range. Interestingly the
$pn$ correlations, which in the ground state
are reduced due to the neutron excess, 
are roughly constant for $T<1$ MeV (in fact they even
slightly increase when the dominating like-nucleon correlations get
weaker). As proposed in Ref. \cite{Engel} in the presence of a neutron excess 
the isovector pairing interaction supports
the separation of the nuclear system at low temperatures into neutron
and proton condensates. At higher temperatures, where the interaction is
less important, the isovector pairing correlations follow the ordering
of the mean-field values and slowly vanish.

\section{Conclusion}

We have studied pairing correlations in self-conjugate
nuclei in the middle of the $pf$ shell
using the pairing plus quadrupole hamiltonian
and the SMMC method.
Several results of our investigation
are noteworthy.

The isovector $J=0$ pairing correlations show a significant staggering
between odd-odd and even-even $N=Z$ nuclei,
as noted previously in calculations based on the realistic
interaction as well as on a schematic analytically solvable
model. While the three isovector
channels have identical strengths in even-even $N=Z$ nuclei, 
the total isovector pairing strength is strongly redistributed
in odd-odd self-conjugate nuclei, with a strong enhancement of the 
proton-neutron correlations. 

The importance of isovector proton-neutron correlations decrease drastically
if neutrons are added, again in accordance with calculations
based on realistic forces and on the analytically solvable model. 
The additional neutrons increase the coherence
among the neutron pair condensate, thus making less neutrons available
for isovector proton-neutron correlations.
At the same time, the correlations among protons also increase
if neutrons are added.

We have studied the temperature dependence of the pairing correlations
and of selected observables for $^{54}$Co and $^{60}$Zn.
The even-even $N=Z$ nucleus $^{60}$Zn shows the same qualitative trends
as other even-even nuclei in this mass region (including 
the pair breaking transition  at temperatures near $T=1$ MeV).
The odd-odd nucleus $^{54}$Co has a different behavior.
While the proton-proton and neutron-neutron correlations (although
much weaker than in even-even nuclei)
show a phase transition near $T=1$ MeV, the dominant 
$J=0$ proton-neutron correlations decrease 
sharply already at lower temperature, and near $T = 1$ MeV
become equal to the like-particle correlations.
We conjecture that the presence of low-lying isospin $T = 0$
states in this odd-odd nucleus is responsible for this behavior.
The temperature dependence of the isospin expectation value
confirms this conjecture. By comparing the isospin expectation values
for the schematic Hamiltonian and for the 
realistic KB3 interaction we conclude that at moderate temperatures 
$pf$ shell nuclei are dominated by isoscalar correlations.
Hence there is a transition from isovector to isoscalar dominance in
odd-odd $N=Z$ $pf$-shell nuclei with increasing temperature.
The isoscalar correlations are the focus of work in progress.

\acknowledgements

Petr Vogel thanks the Danish Research
Council and the Theoretical Astrophysics Center for financial support
during his stay at the University of Aarhus.
This work was supported in part 
by the U.S. Department
of Energy under grant DE-FG03-88ER-40397 and by the NSF grant
PHY94-20470.

\narrowtext
\begin{figure}
\caption{Pairing correlations 
${\cal N}_{t_z}$
in the ground states of the $N=Z$ nuclei
with masses $A=46-74$. 
The full lines show the BCS $pn$ correlations,
while the dotted lines show the $pp$=$nn$ correlations.
In this and following figures the error bars of the SMMC
calculations are indicated. In addition, for $24 \leq N \leq 30$
the results obtained with the realistic KB3 interaction 
[10,18] are also
shown (not connected by lines to distinguish them).}
\label{fig1}
\end{figure}

\narrowtext
\begin{figure}
\caption{ Pairing correlations 
${\cal N}_{t_z}$
in the odd-odd nucleus $^{58}$Cu
as a function of the scaling $\lambda$ of the pairing interaction
constant.
The full lines show the BCS $pn$ correlations,
while the dotted lines show the $pp$=$nn$ correlations.}
\label{fig2}
\end{figure}

\narrowtext
\begin{figure}
\caption{ Dependence of the expectation value $\langle T^2 \rangle$
(full line) and $\langle J^2 \rangle$ (dashed line) for $^{58}$Cu
on the scaling $\lambda$ of the pairing interaction constant.
(Only few typical error bars of the quantity 
$\langle J^2 \rangle$ are shown for clarity.)}
\label{fig3}
\end{figure}

\narrowtext
\begin{figure}
\caption{ Ground state energies of the $N=Z$ nuclei.}
\label{fig4}
\end{figure}

\narrowtext
\begin{figure}
\caption{ Pairing correlations 
${\cal N}_{t_z}$
in the Ni isotopes
as a function of the neutron number $N$. The full line
is for $pn$ pairing, the dotted line is for $pp$
pairing, and the dashed-dotted line is for $nn$ pairing.
The dotted line with full square points shows
the average pairing $(pp+nn+pn)/3$.}
\label{fig5}
\end{figure}

\narrowtext
\begin{figure}
\caption{ Expectation value $\langle H \rangle$
for $^{54}$Co (full line) and $^{60}$Zn (dotted line) as a function
of temperature.}
\label{fig6}
\end{figure}

\narrowtext
\begin{figure}
\caption{ Pairing correlations 
${\cal N}_{t_z}$
in $^{54}$Co (part(a), full line for
$pn$, dotted line for $pp = nn$) and for $^{60}$Zn (part(b),
full line for $pn$, dotted line for $pp = nn$) as a function of 
temperature. In both panels the line without points indicates
the pairing corresponding to mean field. In that case the $pp$ and $pn$
pairing is essentially identical, and thus only one curve is shown.}
\label{fig7}
\end{figure}

\narrowtext
\begin{figure}
\caption{ Expectation value of isospin, $\langle T^2 \rangle$,
for $^{54}$Co (full line) and $^{60}$Zn (dotted line) as a function
of temperature. The dashed (for $^{54}$Co) 
and dot-dashed (for $^{60}$Zn) lines without points
indicate the $\langle T^2 \rangle$ values corresponding to mean field.
The $\langle T^2 \rangle$ values calculated with the KB3 interaction
for temperatures $T >  2.5$ MeV are shown as dotted lines with points;
the higher one corresponding to $^{54}$Co.}
\label{fig8}
\end{figure}

\narrowtext
\begin{figure}
\caption{ Expectation value of angular momentum, $\langle J^2 \rangle$,
for $^{54}$Co (full line) and $^{60}$Zn (dotted line) as a function
of temperature. The mean field values are shown as the 
long-dashed line for $^{54}$Co and the double-dotted line for
$^{60}$Zn.}
\label{fig9}
\end{figure}

\narrowtext
\begin{figure}
\caption{ Expectation value $\langle H \rangle$
for $^{60}$Ni as a function
of temperature.}
\label{fig10}
\end{figure}

\narrowtext
\begin{figure}
\caption{ Pairing correlations 
${\cal N}_{t_z}$
in $^{60}$Ni as a function of
temperature. The full line shows the $pn$ correlations, the dotted
line the $pp$ correlations, and the dot-dashed line the $nn$
correlations.}
\label{fig11}
\end{figure}

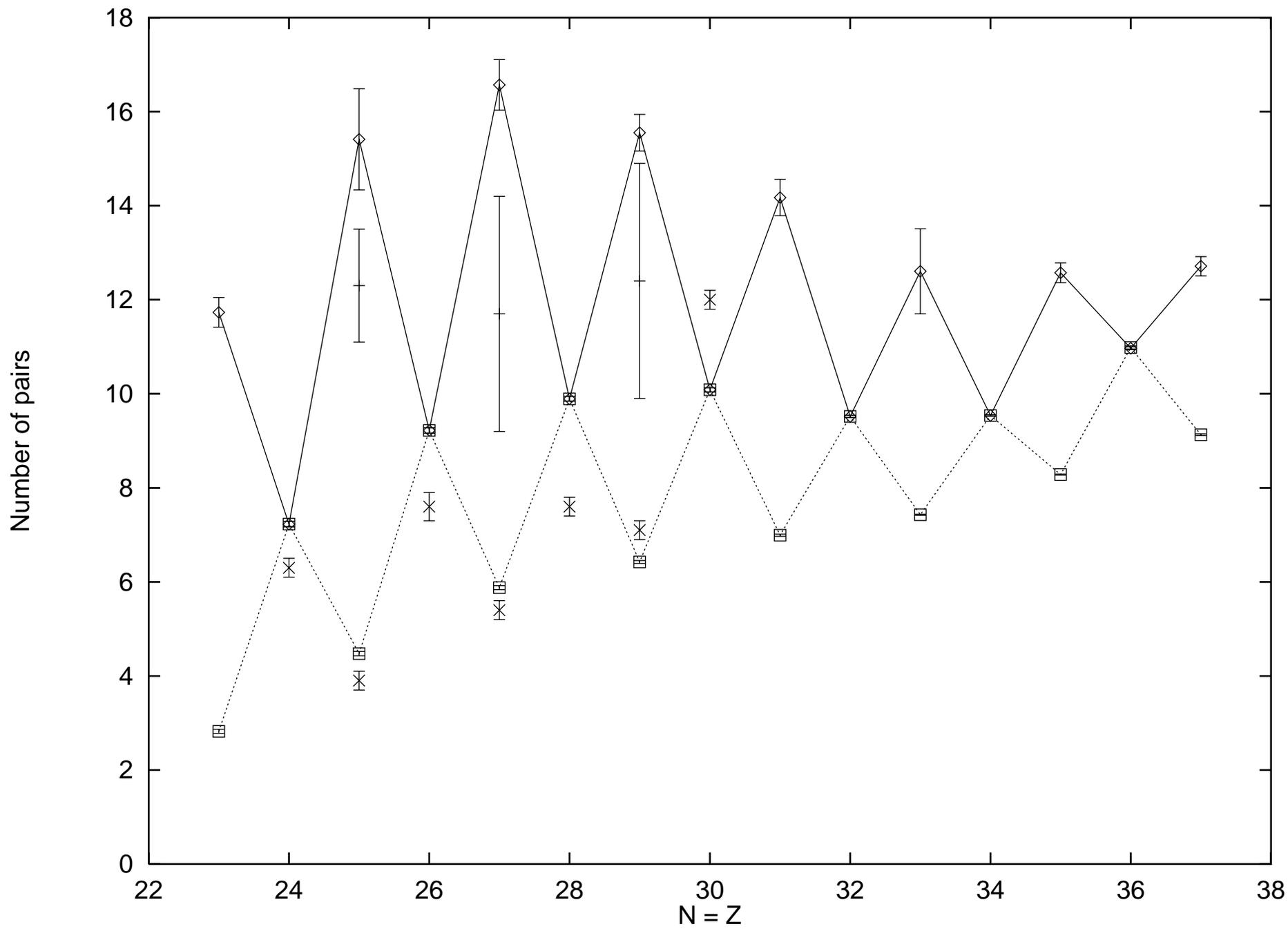
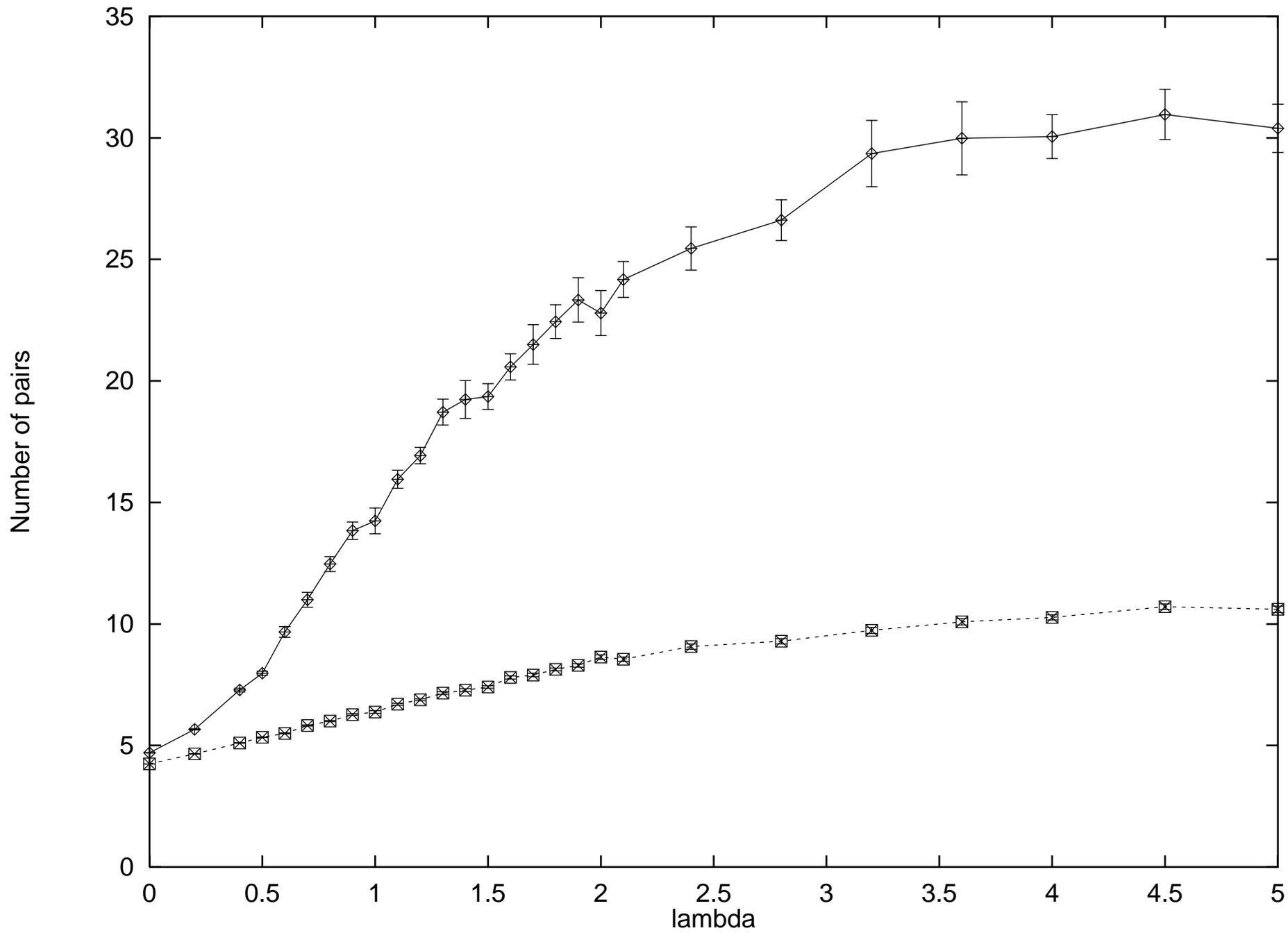
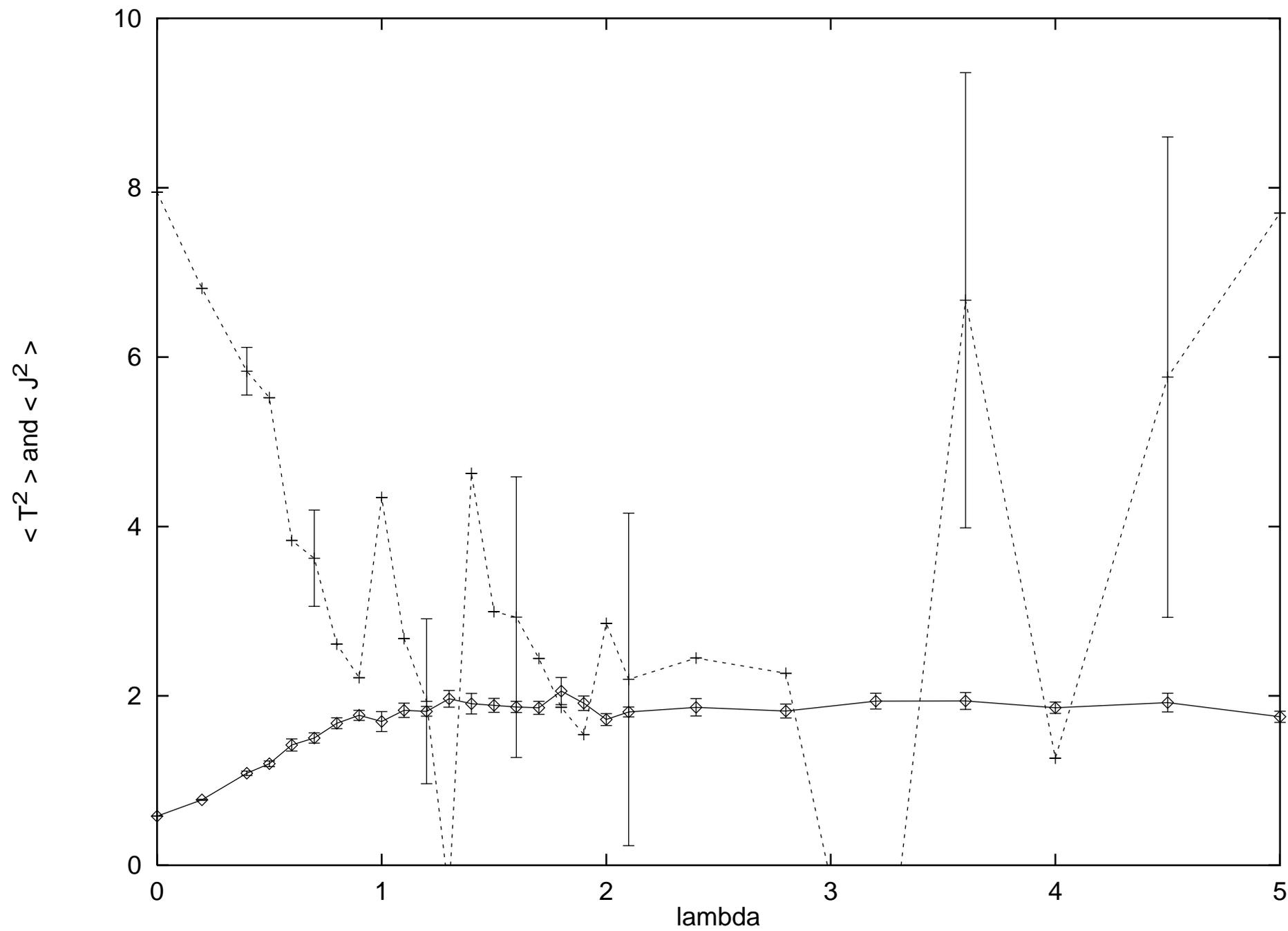
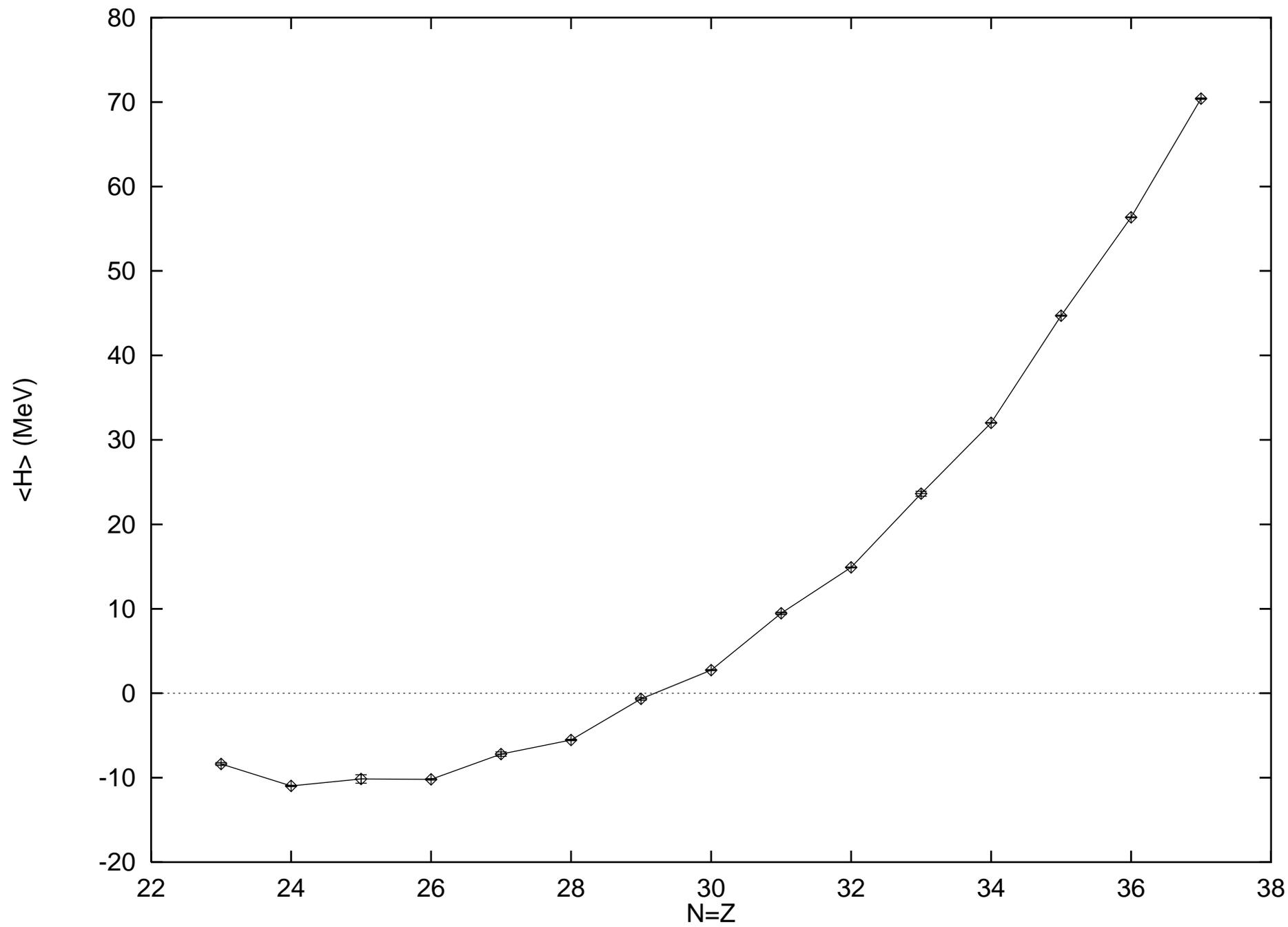
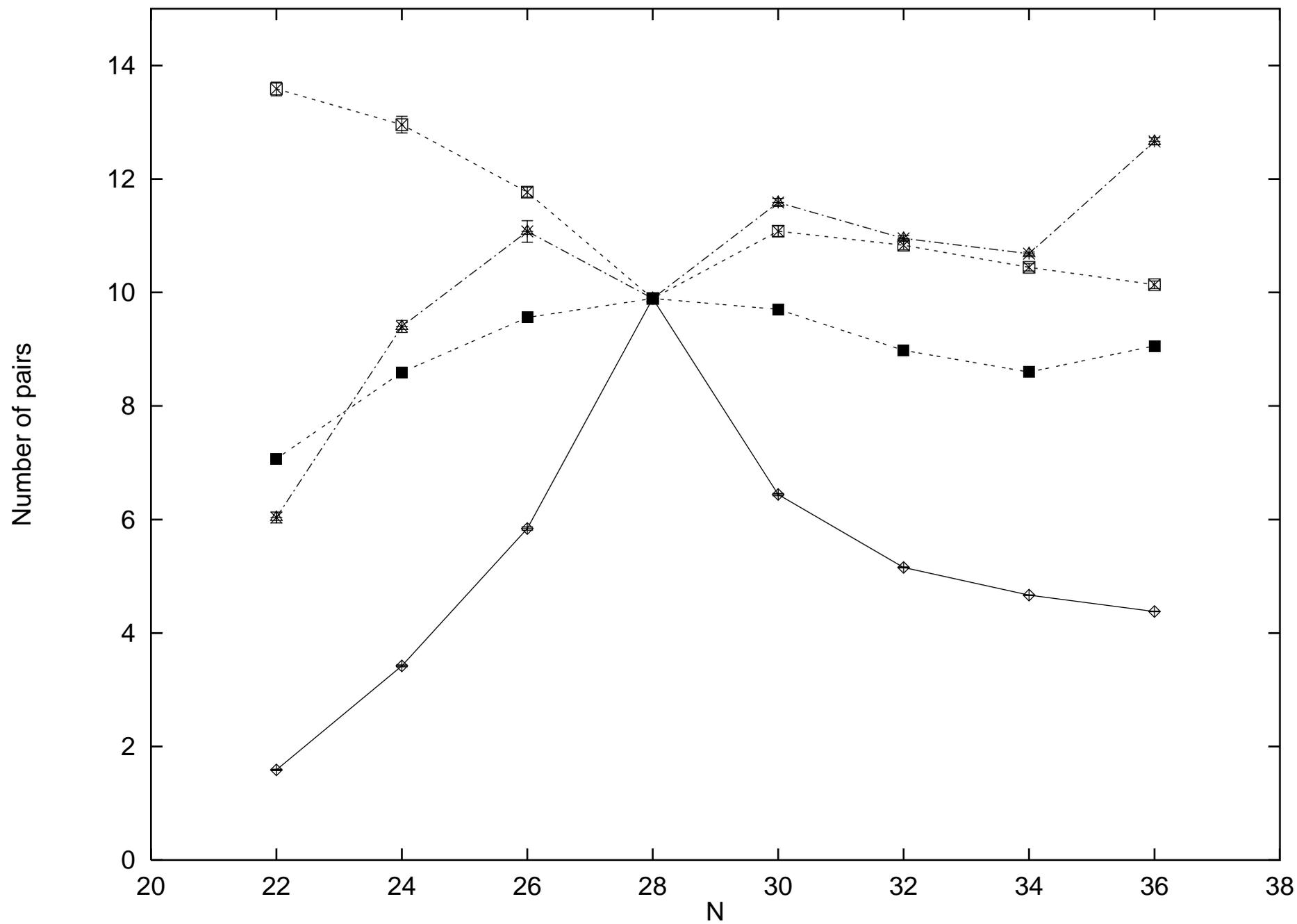
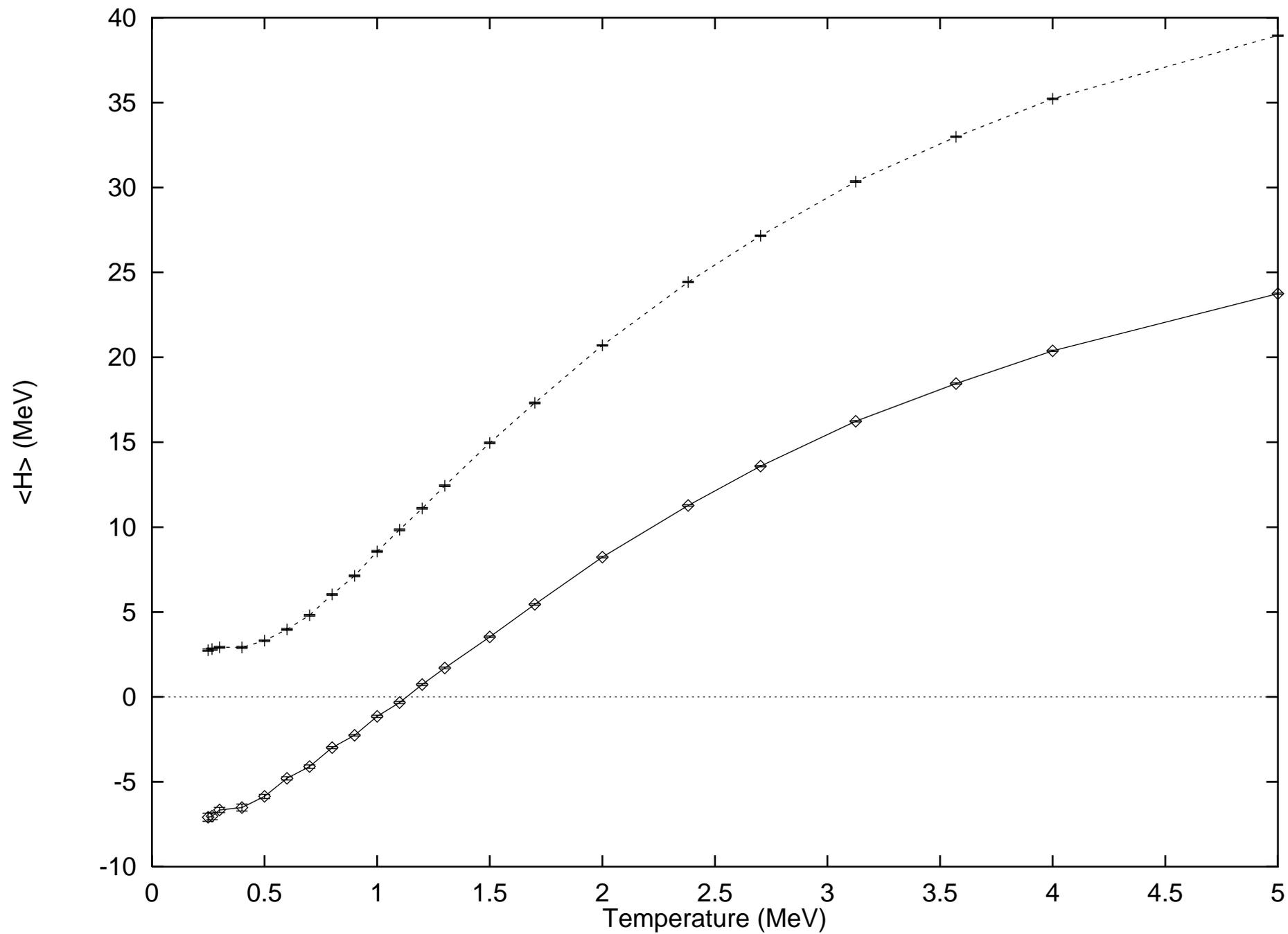
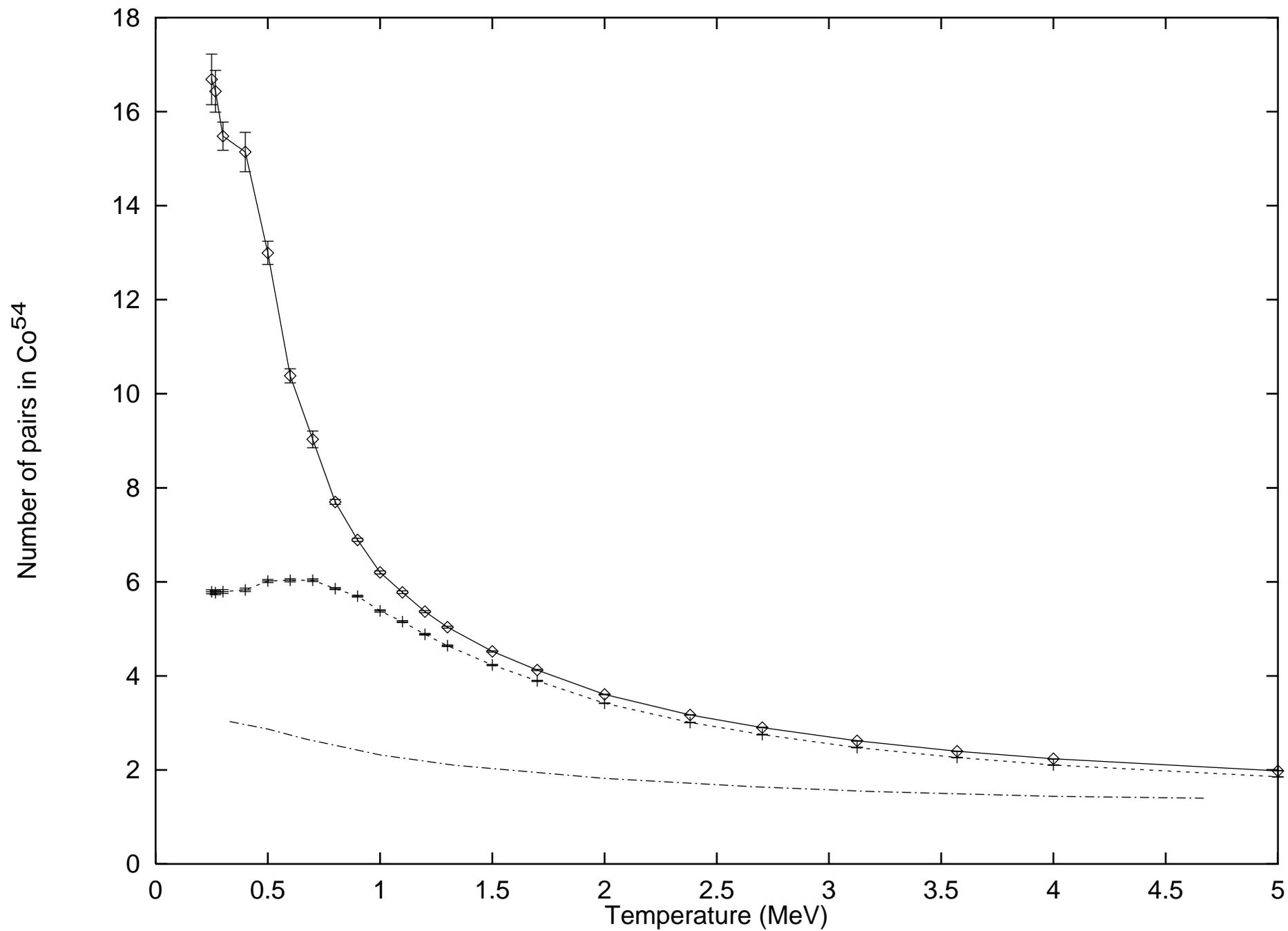
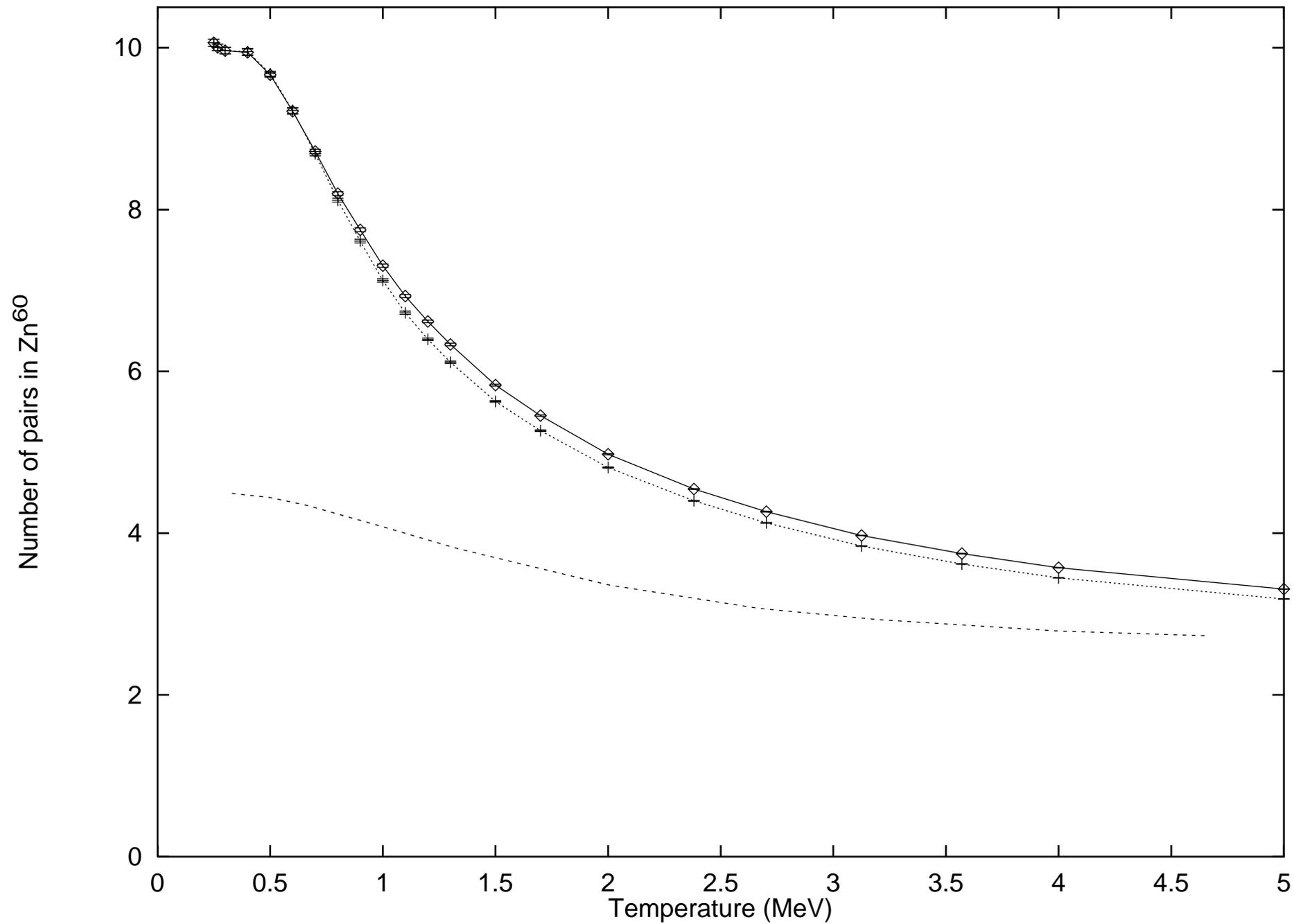
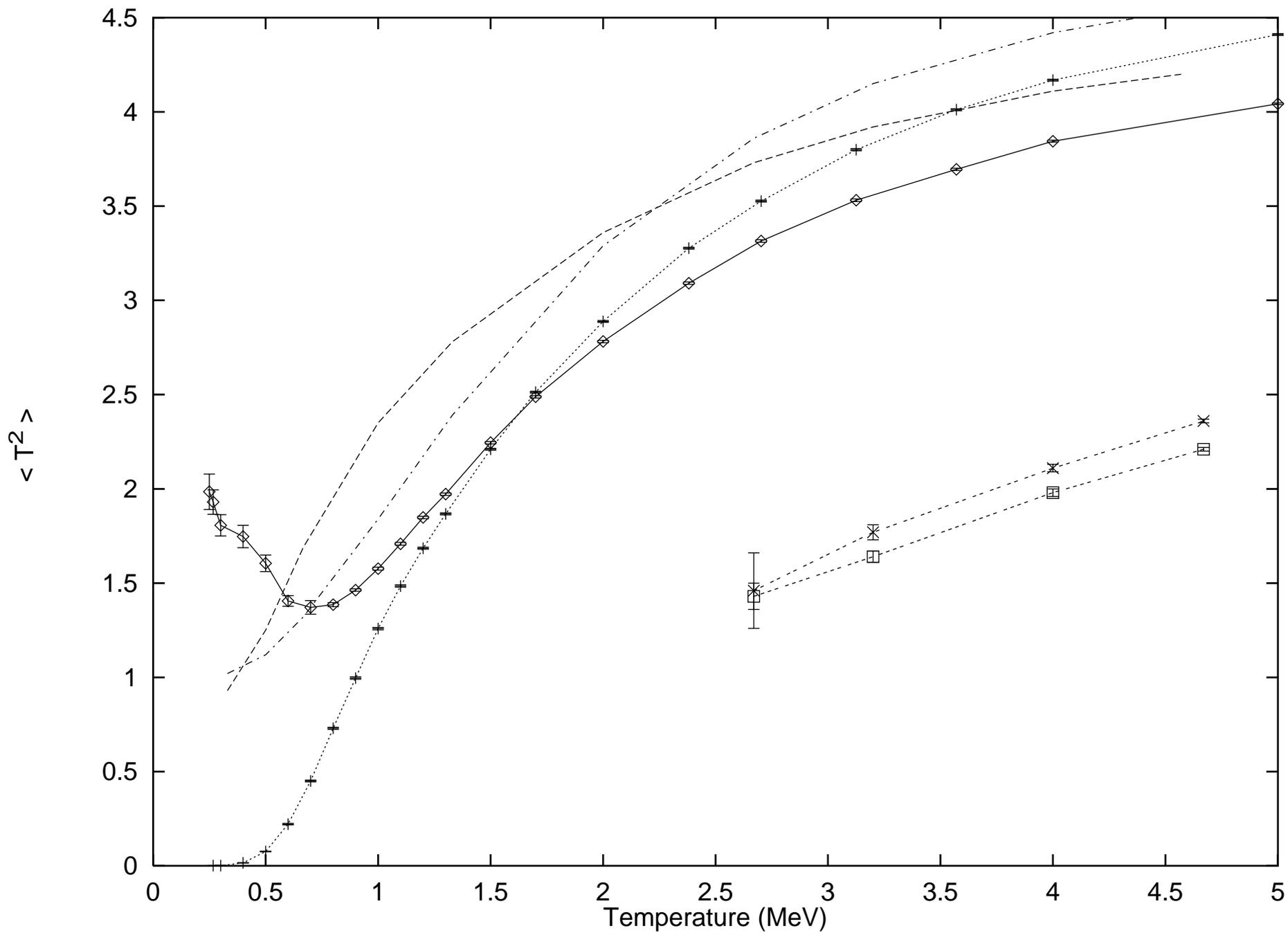
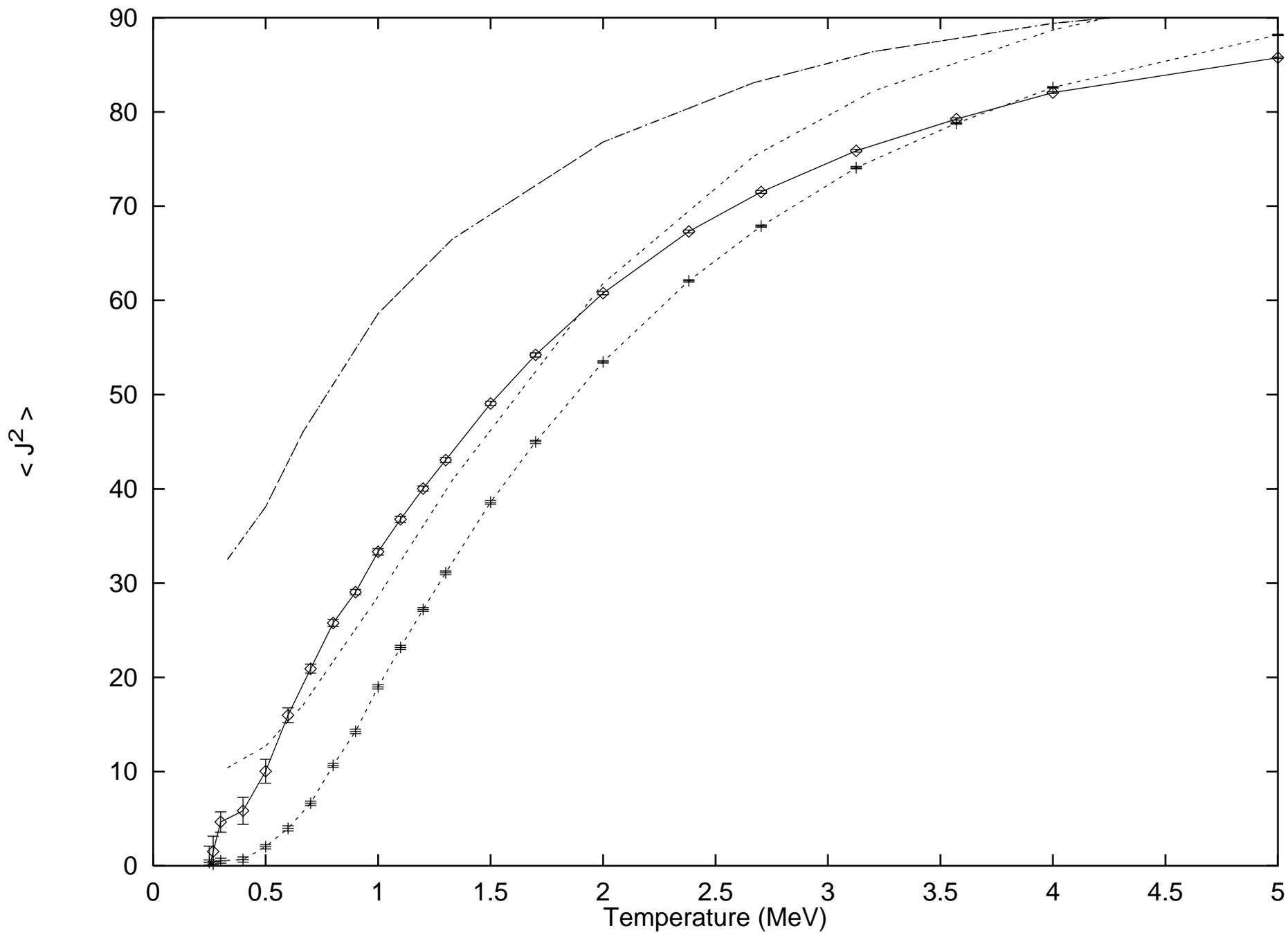
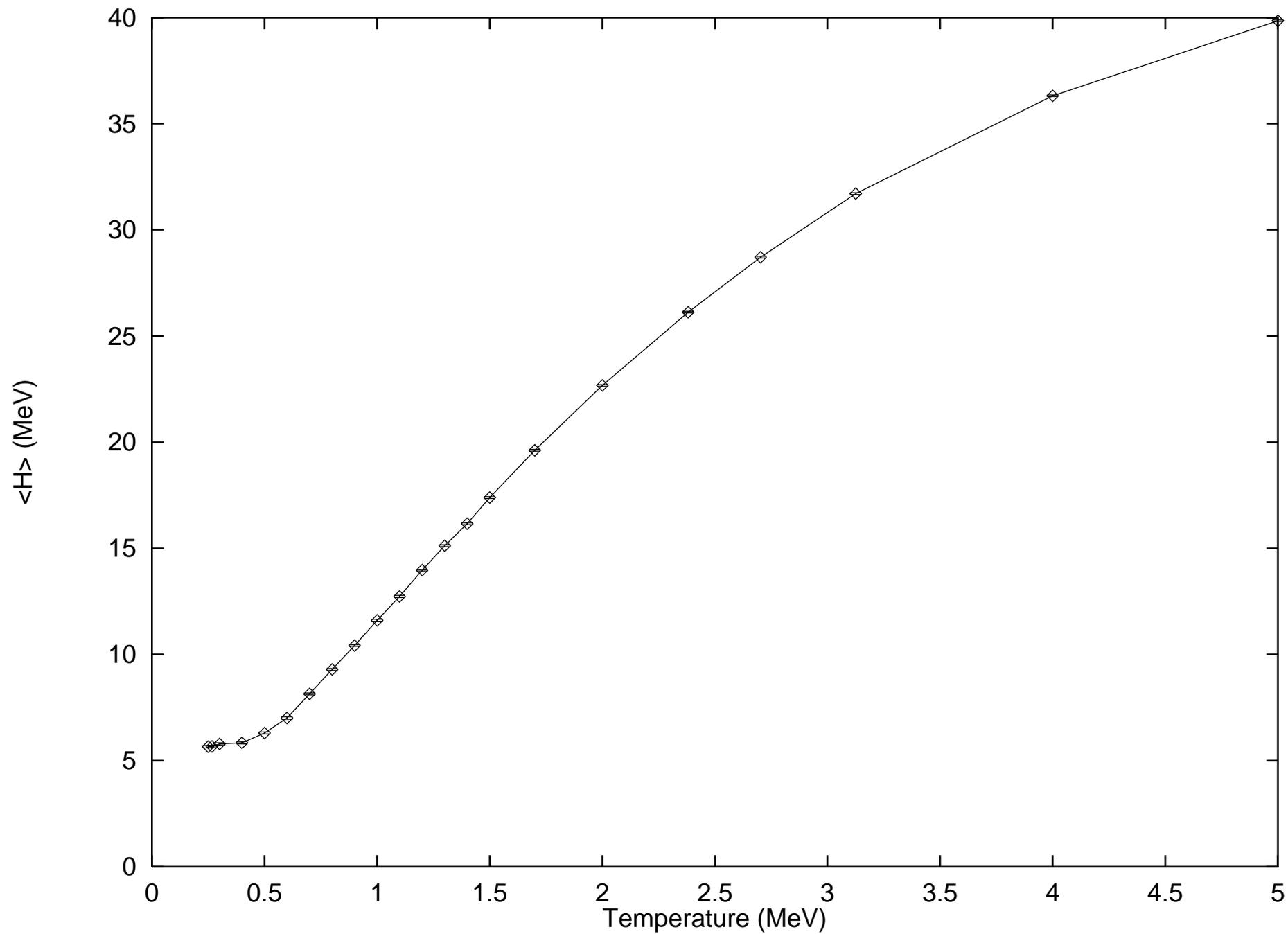
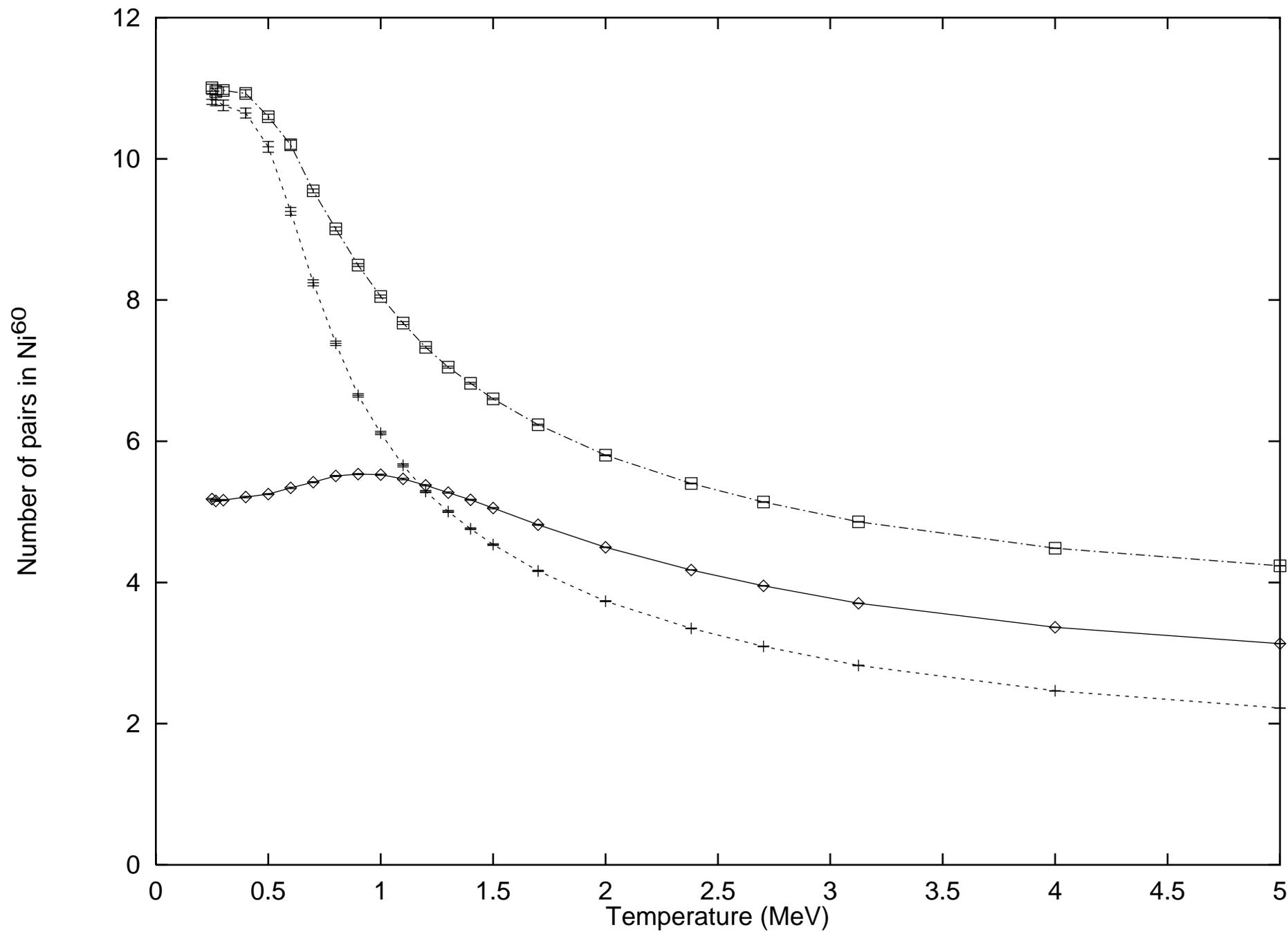
\end{document}